\documentclass[epj]{svjour}
\usepackage{graphics}
\usepackage{amssymb}
\begin{document}
\title{The $\Phi\to K^0\bar K^0\gamma$ decay}
\author{Rafel Escribano\inst{1}
\thanks{\emph{Acknowledgements:}
This work was supported in part by the Ramon y Cajal program,
the Ministerio de Educaci\'on y Ciencia under grant FPA2005-02211,
the EU Contract No.~MRTN-CT-2006-035482, ``FLAVIAnet'', and
the Generalitat de Catalunya under grant 2005-SGR-00994.}%
}                     
%
%
\institute{
Grup de F\'{\i}sica Te\`orica and IFAE,
Universitat Aut\`onoma de Barcelona,
E-08193 Bellaterra (Barcelona), Spain
}
%
\date{
}
%
\abstract{
The scalar contributions to the radiative decay $\phi\to K^0\bar K^0\gamma$
are studied within the framework of the Linear Sigma Model.
Theoretical predictions for the associated subprocesses
$\phi\to f_0\gamma$ and $\phi\to a_0\gamma$
as well as the ratio $\phi\to f_0\gamma/a_0\gamma$ are also given.
\PACS{
      {12.39.-x}{} \and {11.30.Rd}{} \and {13.40.Hq}{} \and {13.75.Lb}{}
     } 
} 
\maketitle
\section{Introduction}
\label{intro}
The radiative decays of the light vector mesons $(V=\rho,\omega,\phi)$
into a pair of neutral pseudoscalars $(P=\pi^0,K^0,\eta)$,
$V\rightarrow P^0P^0\gamma$, are an excellent laboratory for investigating the
nature and extracting the properties of the light scalar meson resonances
$(S=\sigma,a_{0},f_{0})$ \cite{Escribano:2006mb}.
In addition, their study also complements other analysis based on central production,
$D$ and $J/\psi$ decays, etc.~\cite{Close:2002zu}.
Renewed interest on these radiative decays from both the theoretical
and experimental sides are found in Refs.~\cite{Isidori:2006we,Black:2006mn,Achasov:2005hm} and
Ref.~\cite{Ambrosino:2006gk}, respectively.
Particularly interesting is the $\phi\to K^0\bar K^0\gamma$ decay, which, as we will see,
can provide us with valuable information on the properties of the
$f_{0}(980)$ and $a_{0}(980)$ resonances.
The ratio $\phi\rightarrow f_{0}\gamma/a_{0}\gamma$ can also be used to 
extract relevant information on the scalar mixing angle.
At present, there is not yet experimental data available for the $\phi\to K^0\bar K^0\gamma$ process,
while for the ratio $\phi\rightarrow f_{0}\gamma/a_{0}\gamma$, the 
experimental value measured by the KLOE Collaboration is
$R(\phi\rightarrow f_{0}\gamma/a_{0}\gamma)=6.1\pm 0.6$ \cite{Aloisio:2002bs}.

The purpose of this contribution is to compute the $\phi\to K^0\bar K^0\gamma$ decay
where the scalar effects are known to be dominant.
This process is interesting to study, on one side,
because it allows for a direct measurement of the $K\bar K$ couplings to the $f_0$ and $a_0$  mesons thus avoiding a model dependent extraction
and, on the other side, since it could pose a background problem for testing
CP-violation at DA$\Phi$NE.
The direct measurement of the couplings seems to be feasible in the near future with the
higher luminosity expected at DAFNE-2.
Having 50 fb$^{-1}$, the number of expected $K^0\bar K^0\gamma$ final state is in the
range $2\div 8\times 10^3$ \cite{Ambrosino:2006gk}.
The analysis of CP-violating decays in $\phi\rightarrow K^0\bar K^0$ has been proposed
as a way to measure the ratio $\epsilon^\prime/\epsilon$ \cite{Dunietz:1986jf}, 
but because this means looking for a very small effect,
a $B(\phi\rightarrow K^0\bar K^0\gamma)\gtrsim 10^{-6}$
will limit the precision of such a measurement.
Related to the $\phi\to K^0\bar K^0\gamma$ decay,
there are also the processes $\phi\to (f_0,a_0)\gamma$ which are the main contributions
to the former through the decay chain $\phi\to(f_0+a_0)\gamma\to K^0\bar K^0\gamma$.
An accurate measurement of the production branching ratio and of the mass spectra for
$\phi\to f_0(980)/a_0(980)\gamma$  decays can clarify the controversial nature of these
well established scalar mesons.

In Sec.~\ref{theory}, we present a short review of the approaches used in the literature to study
these processes emphasizing the different treatments of the scalar contribution.
The $\phi\to K^0\bar K^0\gamma$ decay is discussed in Sec.~ \ref{K0K0bar}.
The subprocesses $\phi\to f_0\gamma$ and $\phi\to a_0\gamma$
as well as the ratio $\phi\to f_0\gamma/a_0\gamma$ are discussed in Sec.~\ref{ratio}.
Concluding remarks are presented in Sec.~\ref{conclusions}.

\section{Theoretical framework}
\label{theory}
An early attempt to explain the $V\rightarrow P^0P^0\gamma$ decays was done in
Ref.~\cite{Bramon:1992kr} using the vector meson dominance (VMD) model.
In this framework, the $V\rightarrow P^0P^0\gamma$ decays proceed through the
decay chain $V\rightarrow VP^0\rightarrow P^0P^0\gamma$, where
the intermediate vectors exchanged are $V=\bar K^{\ast 0}$ and 
$V^\prime=K^{\ast 0}$ for $\phi\rightarrow K^0\bar K^0\gamma$.
The calculated branching ratio is found to be
$B_{\phi\rightarrow K^0\bar K^0\gamma}^{\rm VMD}=2.7\times 10^{-12}$
\cite{Bramon:1992kr}.
Later on, the $V\rightarrow P^0P^0\gamma$ decays were studied in a 
Chiral Perturbation Theory (ChPT) context enlarged to included on-shell vector
mesons \cite{Bramon:1992ki}. In this formalism,
$B_{\phi\rightarrow K^0\bar K^0\gamma}^\chi=7.6\times 10^{-9}$.
Taking into account both chiral and VMD contributions, one finally obtains
$B_{\phi\rightarrow K^0\bar K^0\gamma}^{{\rm VMD}+\chi}=7.6\times 10^{-9}$
\cite{Bramon:1992ki}.
Additional contributions are those coming from the exchange of scalar resonances.
A first model including the scalar resonances explicitly is the
\emph{no structure model}, where the $V\rightarrow P^0P^0\gamma$ decays
proceed through the decay chain
$V\rightarrow S\gamma\rightarrow P^0P^0\gamma$
and the coupling $VS\gamma$ is considered as pointlike.
A second model is the \emph{kaon loop model} \cite{Achasov:1987ts},
where the initial vector decays into a pair of charged kaons that,
after the emission of a photon, rescatter into a pair of neutral pseudoscalars
through the exchange of scalar resonances.
In Ref.~\cite{Achasov:1987ts}, it was shown for the first time the convenience of studying the
$\phi\rightarrow K^0\bar K^0\gamma$ process
to investigate the nature of the $f_{0}$ and $a_{0}$ scalar mesons.
In this pioneer work,
$B(\phi\rightarrow (f_0+a_0)\gamma\rightarrow K^0\bar K^0\gamma)=1.3\times 10^{-8}$,
for a four-quark structure of the $f_{0}$ and $a_{0}$, and
$2.0\times 10^{-9}$, for a two-quark structure.
The previous two models include the scalar resonances {\it ad hoc},
and the pseudoscalar rescattering amplitudes are not chiral invariant.
This problem is solved in the next two models which are based not only on the
\emph{kaon loop model} but also on chiral symmetry.
The first one is a chiral unitary approach (U$\chi$) where the 
scalar resonances are generated dynamically by unitarizing the one-loop
pseudoscalar amplitudes.
In this approach,
$B_{\phi\rightarrow K^0\bar K^0\gamma}^{\rm U\chi}=5\times 10^{-8}$
\cite{Oller:1998ia}.
The second model is the Linear Sigma Model (L$\sigma$M), a well-defined
$U(3)\times U(3)$ chiral model which incorporates {\it ab initio} the
pseudoscalar and scalar mesons nonets.
The advantage of the L$\sigma$M is to incorporate explicitly the effects of
scalar meson poles while keeping the correct behaviour at low invariant masses
expected from ChPT.

In the next two sections, we discuss the scalar contributions to the
$\phi\rightarrow K^0\bar K^0\gamma$ decay
and the ratio $\phi\rightarrow f_{0}\gamma/a_{0}\gamma$
in the framework of the L$\sigma$M.

\section{$\phi\rightarrow K^0\bar K^0\gamma$}
\label{K0K0bar}
As stated in the Introduction,
this process is the only radiative decay where the $f_0$ and $a_0$ scalar mesons
contribute simultaneously.
This will allow, once this decay is measured presumably at DAFNE-2,
to extract relevant information on the nature and couplings of both mesons,
and to compare it with the one obtained from the already experimentally measured
$\phi\rightarrow\pi^0\pi^0\gamma$ and $\phi\rightarrow\pi^0\eta\gamma$ decays.
Therefore, a theoretical prediction for the different contributions to the
branching ratio and mass spectrum of this process is welcome and useful.
In addition, our analysis will serve to certify that  the decay $\phi\rightarrow K^0\bar K^0\gamma$
cannot pose a background problem for testing CP-violation at DA$\Phi$NE
since the calculated branching ratio is well below $10^{-6}$ (see below).

The scalar contribution to this process is driven by the decay chain
$\phi\rightarrow K^+K^-(\gamma)\rightarrow K^0\bar K^0\gamma$.
The contribution from pion loops is known to be negligible due to $G$-parity and the Zweig rule.
The amplitude for 
$\phi(q^\ast,\epsilon^\ast)\rightarrow K^0(p)\bar K^0(p^\prime)\gamma(q,\epsilon)$
is given by \cite{Escribano:2006mb}
\begin{equation}
\label{AphiK0K0barLsM}
{\cal A}_{\phi\rightarrow K^0\bar K^0\gamma}^{\mbox{\scriptsize L$\sigma$M}}= 
\frac{eg_{s}}{2\pi^2 m^2_{K^+}}\,\{a\}\,L(s)\times
{\cal A}_{K^+ K^-\rightarrow K^0\bar K^0}^{\mbox{\scriptsize L$\sigma$M}} \ ,
\end{equation}
where
$\{a\}=(\epsilon^\ast\cdot\epsilon)\,(q^\ast\cdot q)-
           (\epsilon^\ast\cdot q)\,(\epsilon\cdot q^\ast)$, 
$m^2_{K^0\bar K^0}\equiv s$ is the dikaon invariant mass and $L(m^2_{K^0\bar K^0})$
is a loop integral function.
The $\phi K\bar K$ coupling constant $g_{s}$ takes the value $|g_s|\simeq 4.5$
to agree with $\Gamma_{\phi\rightarrow K^+K^-}^{\rm exp}= 2.09$ MeV \cite{Eidelman:2004wy}.
The $K^+ K^-\rightarrow K^0\bar K^0$ amplitude in Eq.~(\ref{AphiK0K0barLsM}) is calculated from the
L$\sigma$M at tree level and turns out to be\footnote{
For a detailed derivation of Eq.~(\ref{AKpKmpi0pi0ChPTLsM}) starting from the
$U(3)\times U(3)$ L$\sigma$M, see Ref.~\cite{Escribano:2006mb}.}
\begin{equation}
\label{AKpKmpi0pi0ChPTLsM}
\begin{array}{l}
{\cal A}_{K^+K^-\rightarrow K^0\bar K^0}^{\mbox{\scriptsize L$\sigma$M}}=
\frac{s-m^2_{K}}{4f_K^2}\left[
\frac{m^2_K-m^2_{\sigma}}{D_{\sigma}(s)}({\rm c}\phi_S -\sqrt{2}{\rm s}\phi_S)^2
\right.\\
\qquad\left.
   +\frac{m^2_K-m^2_{f_0}}{D_{f_0}(s)}({\rm s}\phi_S +\sqrt{2}{\rm c}\phi_S)^2
   -\frac{m^2_K-m^2_{a_0}}{D_{a_0}(s)}\right]\nonumber\\
\qquad
+\frac{m^2_{K}-s/2}{2f_K^2}\ ,
\end{array}
\end{equation}
where $D_{S}(s)$ are the $S=\sigma, f_0, a_0$ propagators,
$\phi_S$ is the scalar mixing angle in the quark-flavour basis
and $({\rm c}\phi_S, {\rm s}\phi_S)\equiv (\cos\phi_S, \sin\phi_S)$.
A Breit-Wigner propagator is used for the $\sigma$, while for the $f_{0}$ and $a_0$
a complete one-loop propagator taking into account finite width 
corrections is preferable (see Ref.~\cite{Escribano:2006mb} for details).
A few remarks on the four-pseudoscalar amplitude in Eq.~(\ref{AKpKmpi0pi0ChPTLsM})
are of interest.
First, for $m_S\to\infty$ $(S=\sigma, f_0, a_0)$, the L$\sigma$M amplitude (\ref{AKpKmpi0pi0ChPTLsM})
reduces to the corresponding chiral-loop amplitude $s/4f_K^2$ \cite{Bramon:1992ki},
thus satisfying the chiral constraints and making the whole analysis quite reliable.
Hence, the amplitude (\ref{AKpKmpi0pi0ChPTLsM}) has to be considered as an improved expression of its chiral-loop counterpart mentioned before taking into account the (pole dominated)
$s$-channel scalar dynamics.
Second, the large widths of the scalar resonances break chiral symmetry if they are naively
introduced in amplitudes, an effect already noticed in Ref.~\cite{Achasov:1994iu}.
Accordingly, we introduce the widths in the scalar meson propagators
only {\it after} chiral cancellation of constant terms in the original L$\sigma$M amplitude.
In this way the pseudo-Goldstone nature of kaons is preserved.

\begin{figure}
\centerline{\includegraphics{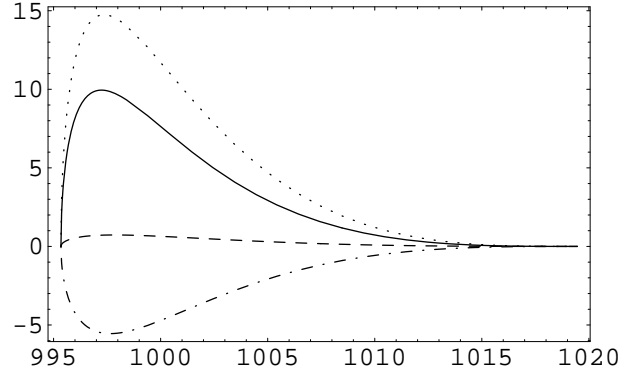}}
\caption{\small
$dB(\phi\rightarrow K^0\bar K^0\gamma)/dm_{K^0\bar K^0}\times 10^9$ (in units of MeV$^{-1}$)   
as a function of the dikaon invariant mass $m_{K^0\bar K^0}$ (in MeV).
The dotted, dashed and dot-dashed lines correspond to the separate contributions 
from $f_0$, $a_0$ and their interference, respectively. 
The solid line is the total result.}
\label{dBdmKKgamma}
\end{figure}

The final results for ${\cal A}(\phi\rightarrow K^0\bar K^0\gamma)$ are then the
sum of the L$\sigma$M contribution in Eq.~(\ref{AphiK0K0barLsM}) plus the VMD
contribution that can be found in Ref.~\cite{Escribano:2006mb}.
The $K^0\bar K^0$ invariant mass distribution,
with the separate contributions from $f_0$, $a_0$ and their interference,  
as well as the total result, are shown in Fig.~\ref{dBdmKKgamma}.
These mass spectra are computed assuming masses for the scalar resonances of
$m_{f_0}=985$ MeV and $m_{a_0}=984.7$ MeV,
and a pseudoscalar (scalar) mixing angle of $\phi_P=41.8^\circ$ ($\phi_S=-8^\circ$).
The values of the $f_0$ mass and the scalar mixing angle are obtained from the
$\phi\rightarrow\pi^0\pi^0\gamma$ analysis in Ref.~\cite{Escribano:2006mb}
whereas the $a_0$ mass is taken from Ref.~\cite{Eidelman:2004wy} and
$\phi_P$ from the ratio $\phi\to\eta^\prime\gamma/\eta\gamma$ \cite{Aloisio:2002vm}.
As seen, the $f_0$ contributes more strongly than the $a_0$
due to a smaller imaginary part of the propagator and a larger coupling to kaons.
The interference is negative since isospin invariance implies
$g_{f_0K^+K^-}=g_{f_0K^0\bar K^0}$ and $g_{a_0K^+K^-}=-g_{a_0K^0\bar K^0}$.
Integrating the $K^0\bar K^0$ invariant mass spectrum one obtains for the scalar contribution
$B(\phi\to K^0\bar K^0\gamma)_{\mbox{\scriptsize L$\sigma$M}}=7.5\times 10^{-8}$.
It is worth mentioning that the value obtained is very sensitive to the scalar mixing angle,
a change of $1^\circ$ modifies the branching ratio around 20\%.
The whole scalar contribution includes not only the $f_0$ and $a_0$ effects but also the $\sigma$ ones.
However, the latter are negligible due to the suppression of the $\sigma K\bar K$ coupling if\footnote{
A value of $m_{\sigma}=478$ MeV is taken from the analysis of $D^+\to\pi^-\pi^+\pi^+$ performed
by the E791 Coll.~\cite{Aitala:2000xu} at Fermilab.}
$m_\sigma\simeq m_K$ and for kinematical reasons.
Numerically, they amount to less than 1\% of the total result.
Finally, the exchange of intermediate $K^{\ast 0}$ and $\bar K^{\ast 0}$
vector mesons is calculated to be
$B(\phi\to K^0\bar K^0\gamma)_{\rm VMD}=2.0\times 10^{-12}$ and therefore negligible.

\section{$\phi\rightarrow (f_{0}, a_{0})\gamma$ and $\phi\rightarrow f_{0}\gamma/a_{0}\gamma$}
\label{ratio}
In the kaon loop model, these two processes are driven by the decay 
chain
$\phi\rightarrow K^+K^-(\gamma)\rightarrow f_{0}\gamma$ and
$a_{0}\gamma$.
The amplitudes are given by
\begin{equation}
\label{Aratio}
{\cal A}=\frac{eg_{s}}{2\pi^2 m^2_{K^+}}\,\{a\}\,L(m^2_{f_{0}(a_{0})})
\times g_{f_{0}(a_{0})K^+K^-}\ ,
\end{equation} 
where the scalar coupling constants are fixed within the L$\sigma$M to
\begin{equation}
\label{couplingsratio}
\begin{array}{c}
g_{f_{0}K^+K^-}=\frac{m^2_{K}-m^2_{f_{0}}}{2f_{K}}
({\rm s}\phi_S +\sqrt{2}{\rm c}\phi_S)\ ,\\[2ex]
g_{a_{0}K^+K^-}=\frac{m^2_{K}-m^2_{a_{0}}}{2f_{K}}\ .
\end{array}
\end{equation}
The ratio of the two branching ratios is thus
\begin{equation}
\label{R}
\begin{array}{rcl}
R_{\phi\rightarrow f_{0}\gamma/a_{0}\gamma}^{\mbox{\scriptsize L$\sigma$M}}
&=&
\frac{|L(m^2_{f_{0}})|^2}{|L(m^2_{a_{0}})|^2}
\frac{\left(1-m^2_{f_{0}}/m^2_{\phi}\right)^3}
       {\left(1-m^2_{a_{0}}/m^2_{\phi}\right)^3}
       \times\frac{g^2_{f_{0}K^+K^-}}{g^2_{a_{0}K^+K^-}}\\[2ex]
&\simeq&
({\rm s}\phi_S +\sqrt{2}{\rm c}\phi_S)^2\ ,
\end{array}
\end{equation}
where the approximation is valid for $m_{f_{0}}\simeq m_{a_{0}}$.
Integrating the corresponding amplitude in Eq.~(\ref{Aratio}) one obtains the branching ratios
$B(\phi\rightarrow f_{0}\gamma)=2.6\times 10^{-4}$ and
$B(\phi\rightarrow a_{0}\gamma)=1.6\times 10^{-4}$, respectively.
For the ratio (\ref{R}) one gets
$R_{\phi\rightarrow f_{0}\gamma/a_{0}\gamma}^{\mbox{\scriptsize L$\sigma$M}} \simeq 1.6$
for $\phi_{S}=-8^\circ$.
These predictions should be compared with the experimental measurements
$B(\phi\rightarrow f_{0}\gamma)=(4.40\pm 0.21)\times 10^{-4}$,
$B(\phi\rightarrow a_{0}\gamma)=(7.6\pm 0.6)\times 10^{-5}$ \cite{Eidelman:2004wy} and
$R_{\phi\rightarrow f_{0}\gamma/a_{0}\gamma}^{\mbox{\scriptsize KLOE}}
 =6.1\pm 0.6$ \cite{Aloisio:2002bs}.
As seen, the discord is clear.
However, the first value, which is based on the KLOE study of the $\phi\to\pi^0\pi^0\gamma$ decay
 \cite{Aloisio:2002bt}, is obtained from a large destructive interference 
between the $f_{0}\gamma$ and $\sigma\gamma$ contributions to
$\phi\rightarrow\pi^0\pi^0\gamma$, in disagreement with other 
experiments \cite{Achasov:2000ym}.
Consequently, the experimental value for the ratio (\ref{R}) could be overestimated.
In addition, the approximate expression for this ratio is only valid when the $f_0$ mass
is below the charged kaon threshold ($2m_{K^+}\simeq 987$ MeV).
If not, the steep behaviour of the loop function after threshold makes the approximation meaningless
and the exact expression has to be taken.
In this case, $R_{\phi\rightarrow f_{0}\gamma/a_{0}\gamma}^{\mbox{\scriptsize L$\sigma$M}}$
can be much smaller than the given prediction (independently of the mixing angle value).
Conversely, if the $a_0$ mass, which we have kept fixed to $m_{a_0}=984.7$ MeV,
is moved to a value above threshold, then the ratio can be even larger than the experimental result.
In this case, our prediction for $B(\phi\rightarrow a_{0}\gamma)$ decreases
and agrees with the experimental result for $m_{a_0}\simeq 989$ MeV.
Therefore, a confirmation of the
$\phi\rightarrow f_{0}\gamma$ and $\phi\rightarrow a_{0}\gamma$ measurements from the analysis
of the $\phi\to\pi^0\pi^0\gamma$ and $\pi^0\eta\gamma$ decays together with a precise fit of the
$f_0$ and $a_0$ masses using these processes is mandatory
before drawing definite conclusions on the validity of our predictions.
One should keep in mind, however, that the experimental data on
$\phi\to\pi^0\pi^0\gamma$ and $\pi^0\eta\gamma$ are satisfactorily accommodated in our framework
(see Ref.~\cite{Escribano:2006mb} for comparison).

\section{Conclusions}
\label{conclusions}
The radiative decay $\phi\rightarrow K^0\bar K^0\gamma$
has been shown to be very useful to extract relevant information on the
properties of the $f_{0}(980)$ and $a_{0}(980)$ scalar mesons.
Our predicted branching ratio including these scalar resonances explicitly is
$B(\phi\to K^0\bar K^0\gamma)=7.5\times 10^{-8}$.
This value is in agreement with previous phenomenological
estimates \cite{Oller:1998ia,Close:ay,Achasov:2001rn}
(see also Ref.~\cite{Close:ay} for a review of earlier predictions).
Notice that the branching ratio obtained here is one order of magnitude larger than the
chiral-loop prediction $B(\phi\to K^0\bar K^0\gamma)=4.1\times 10^{-9}$.
However, it is still one order of magnitude smaller than the limit, ${\cal O}(10^{-6})$,
in order to pose a background problem for testing CP-violating decays at DA$\Phi$NE.
A measurement of this process at DAFNE-2 would be welcome and can serve
as an additional test of the whole approach.
We have also shown that the ratio $\phi\rightarrow f_{0}\gamma/a_{0}\gamma$ can be used to 
obtain valuable information on the scalar mixing angle and on the 
nature of the $f_{0}(980)$ and $a_{0}(980)$ scalar states.

%
%

\end{document}